# Taming the Virtual Space for Incremental Full Configuration Interaction

Jeffrey Hatch, Paul M. Zimmerman
Department of Chemistry, University of Michigan
930 N. University Ave, Ann Arbor, MI 48109, USA, *paulzim@umich.edu

## Abstract

Incremental full configuration interaction (iFCI) closely approximates the FCI limit with polynomial cost through a many-body expansion of the correlation energy, providing highly accurate total energies within a given basis set. To extend iFCI beyond previous basis set limitations, this work introduces a novel natural orbital screening approach, iNO-FCI. By consideration of the importance of virtual orbital selection in the convergence of iFCI, iNO-FCI maximizes the consistency between orbitals selected for each correlated body. iNO-FCI employs a principle of cancellation of errors and ensures that the same set of virtual NOs are used for interdependent terms. This strategy significantly reduces computational cost without compromises in precision. Computational savings of up to 95% are demonstrated, allowing access to larger basis sets that were previously computationally prohibitive. iNO-FCI is herein introduced and benchmarked for several difficult test cases involving double-bond dissociation, biradical systems, conjugated π systems, and the spin gap of a Cu-based transition metal complex.

## Introduction

Studies of electronic structure theory over the last century have elucidated a myriad of chemical phenomena.[1–9] The methods encompassed within electronic structure theory range from tools with modest accuracy to techniques that precisely quantify almost any chemical property.[10–17] Wave function methods in particular are powerful because they can be systematically improved to a desired level of accuracy.[18–20] One of the foundational wave function methods is configuration interaction (CI), which is conceptually the simplest post Hartree-Fock method but also among the most computationally intensive.[21–26] By considering the interaction of a complete set of Slater determinants, strong and weak correlation can be elucidated and the exact energy of the system can be determined. Except for full configuration interaction (FCI), which is exact for any basis choice, the choice of molecular orbitals can have a strong impact on the result of a CI computation. Therefore, the specific means for construction of molecular orbitals must be addressed in any truncated CI method. [27–34]

While FCI is impractical for all but the smallest chemical systems, [35–38] novel variants of CI have permitted accurate wavefunction simulations to be performed on larger systems than ever thought possible. Select CI (SCI) methods follow the same general procedure as FCI but drop the "deadwood" that has negligible impact on the wavefunction. SCI methods therefore have a fraction of the cost of FCI, yet still routinely reach chemical accuracy.[39–46] In SCI methods, the convergence patterns depend significantly on orbital choice, where poor orbitals (e.g., Canonical Hartree Fock orbitals) can lead to slow convergence. Studies have investigated the optimal basis representation for SCI methods and found that natural orbitals (NOs) show improvement over Hartree-Fock orbitals. NOs are known to be similar to—but not the same as—optimized orbitals, where the latter come with substantially increased computational cost.[47–49]

Incremental FCI (iFCI) follows the same spirit as SCI but uses a unique strategy to avoid deadwood in a wavefunction. To achieve this, iFCI utilizes a many-body expansion to represent the wave function in terms of a set of independent bodies, each of which contributes to the FCI limit.[50–52] iFCI has been shown in Refs 51-53 to be effective in approximating FCI results, especially when the bodies of the expansion have certain properties. First, electronic correlation must be treated as a sum over contributions from localized bodies, so iFCI uses a set of localized molecular orbitals as the basic unit for expansion. Second, the Summation NO (SNO) procedure of iFCI compresses the virtual space by removing virtual orbitals that contribute little-to-zero to the correlation energy (see ref 39). Related electronic structure methods, specifically the frozen natural orbital method, employ similar techniques to reduce computational cost by screening virtual orbitals.[53–55]

Based on the above considerations, the orbitals for iFCI are generated through a series of electronic structure calculations. These calculations localize the occupied orbitals and screen the virtual orbital space. Natural orbitals (NOs) are particularly well-suited for iFCI because they facilitate convergence at a faster rate than canonical orbitals and support the screening of virtual orbitals.[56] The proposed iNO-FCI methodology, detailed in the theory section below, allows the iFCI energy to converge towards the FCI energy by the 3- or 4-body expansion level. Traditionally, achieving such convergence with the iFCI method requires using tight screening parameters to determine which virtual orbitals are included in the expansion. Consequently, a large number of orbitals remain in the virtual space for each term of the iFCI expansion. As a result, the primary limitation of iFCI in practice has been the difficulty of scaling to larger basis sets.

The theory section will show how the many-body expansion fundamentally relies on cancellation of redundant terms. For example, four electrons will have a two-body correlation energy that avoids double-counting via subtraction of the two-electron, one-body terms in iFCI. In the limit of infinitely separated electron pairs, the two-body terms should cancel precisely with the two one-body terms, giving zero correlation energy. To achieve this cancellation in practice, the virtual spaces for the two-body term and the two one-body terms should closely align. One way of achieving this is by tightening the SNO screening threshold, with concomitant increase in computational cost. On the other hand, it is conceivable to build this cancellation more deeply into the iFCI procedure, particularly by exercising better control over the composition of the virtual spaces. Herein, we introduce an alternative NO procedure (Figure 1), denoted the incremental NO (iNO) approach, that provides better convergence properties than the original SNO procedure. This work is motivated by but distinct from a prior effort by our group to use natural orbitals in a systematically convergent framework for approximating Hamiltonian eigenvalues.[57]

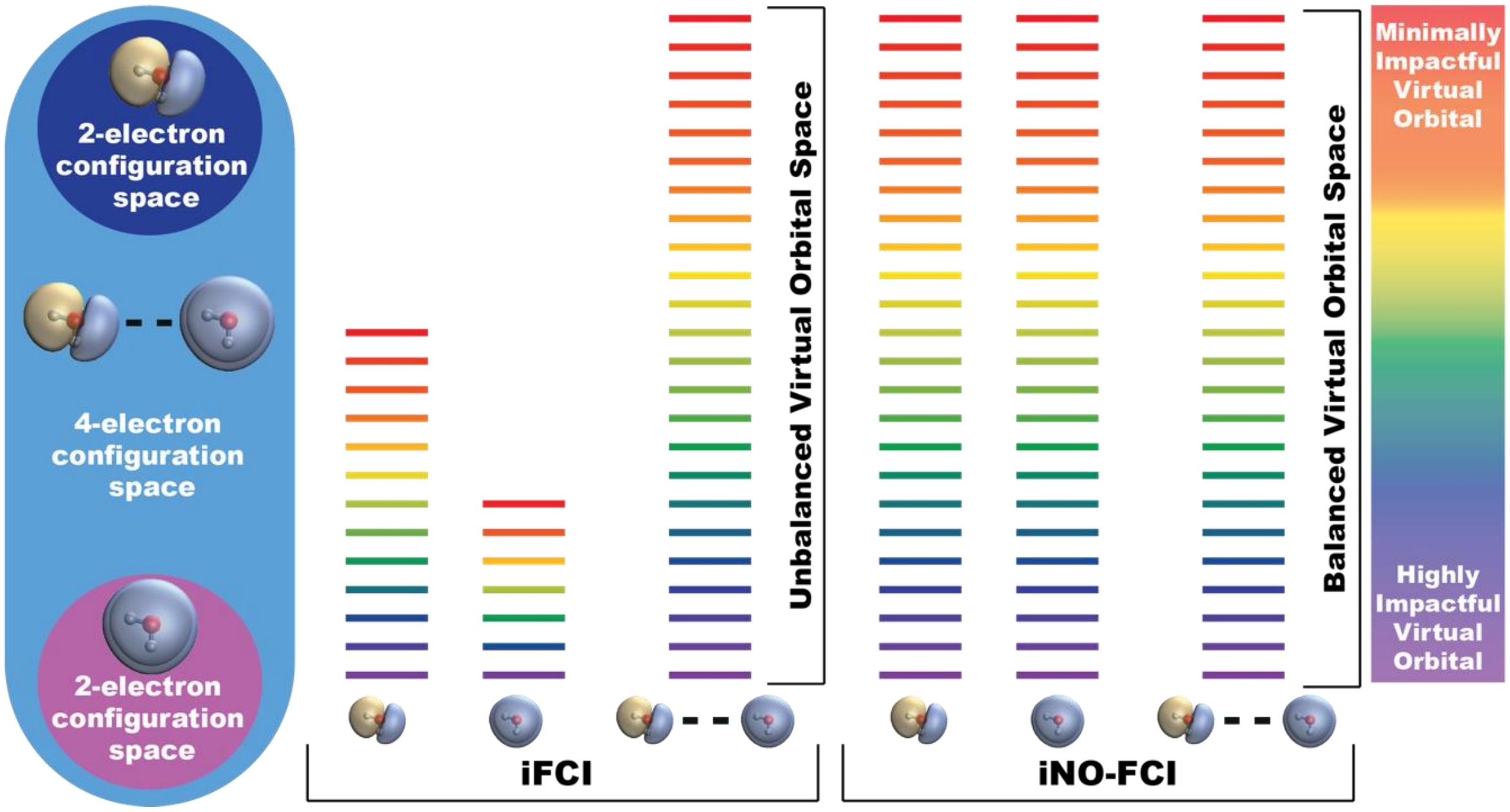

**Figure 1.** iNO ensures perfect alignment of the virtual spaces of interdependent terms whose configuration spaces overlap, allowing for more liberal screening of virtual orbitals without impacting accuracy.

This study aims to expand the applicability of iFCI by leveraging the iNO approach to converge correlated wavefunctions with larger basis sets.[58] The iNO method enables more efficient virtual orbital screening compared to SNO, significantly reducing the computational cost of iFCI calculations.[59] The improved virtual orbital screening and convergence will be demonstrated using a series of tests on challenging cases involving molecules and a transition metal complex. Within the test cases, basis sets as large as polarized quadruple zeta will be usable within iFCI, exceeding what could be done with prior FCI-level computations. The iNO method will be able to resolve electronic states in notoriously difficult systems, including spin gaps, bond dissociation profiles, and reactions of strongly correlated species.[60–65] The largest test case, a Cu-based transition metal complex, involves correlating 130 active electrons in over 400 orbitals.

## **Methods:**

### *Incremental Many-Body Expansion of iFCI*

iFCI is initiated from a set of orthogonal molecular orbitals that represent valence electron pairs. The individual bodies of the expansion are defined to be bonding-antibonding pairs of orbitals, each pair with two electrons. Starting from this reference, the $n$-body expansion treats orbital pairs of size $n$, where $2n$ electrons will be correlated at each level. This decomposes the FCI problem into a manageable series of calculations of polynomial-cost, while also giving a size extensive description of the total energy. The iFCI energy is expressed as

$$E = E_{ref} + \sum_{i} \epsilon_i + \sum_{i<j} \epsilon_{ij} + \sum_{i<j<k} \epsilon_{ijk} + \cdots \quad (1)$$

where

$$\epsilon_i = E_c(i)\,\big|_{\zeta_i} \quad (2)$$

$$\epsilon_{ij} = E_c(ij) - E_c(i) - E_c(j)\,\big|_{\zeta_{ij}} \quad (3)$$

$$\epsilon_{ijk} = E_c(ijk) - E_c(ij) - E_c(ik) - E_c(jk) - E_c(i) - E_c(j) - E_c(k)\,\big|_{\zeta_{ijk}} \quad (4)$$

and the indices, i, j, k… refer to the bodies of the expansion and $E_c(X)$ refers to the correlation energy coming from a CI calculation for bodies $X$. For iFCI with the iNO setup, each correlation energy, $E_c(X')$ is evaluated using the NOs denoted by $\zeta_x$.

Taking as an example the 2-body terms, a particular feature of the iFCI expansion becomes apparent. At the 2-body level, $E_c(ij)$ energies contain the $E_c(i)$ and $E_c(j)$ correlation energies, plus the interactions between the two bodies. $\epsilon_{ij}$ therefore, includes a subtraction of its two 1-body elements to avoid double counting. Furthermore, accurate calculation of terms like $\epsilon_{ij}$ and $\epsilon_{ijk}$ is critical to convergence of iFCI, but remarkably, accurate calculation of $E_c(X)$ is less critical. In iFCI, the accuracy of the $E_c(X)$ terms is dictated by the truncation of the NO space, with NO threshold of $\zeta$. iNO-FCI therefore ensures the virtual spaces of all interdependent terms are the *same*, by using the NOs of a given $n$-body term to recompute all subtractive, lower order terms. In practice, $n = 1$ iNO-FCI is the same as the previous iFCI method, but the $n > 1$ terms are distinct.

*The iterative natural orbital (iNO) approach*

Prior to invoking the iFCI expansion, perfect pairing (PP) molecular orbitals are constructed to localize bonding-antibonding orbital pairs, capturing some static correlation in the reference state (cf. Equation 1) before the n-body expansion begins.[66–69] The initial set of virtual orbitals for each incremental computation is further refined using natural orbitals (NOs) from a low-cost, approximate CI calculation.[70–72] These NOs, denoted as $\zeta_x$, facilitate faster convergence of the CI correlation energy by reducing the number of virtual orbitals needed while still recovering most of the correlation energy. Furthermore, the iNO approach recalculates lower-order terms using these NOs, which mitigates dependence on the size of the virtual space through cancellation of errors between $\epsilon_X$ values. This refinement makes larger basis sets more practical with the iNO-FCI method. Additional details about the iFCI procedure can be found in refs. 50–53 and the Computational Details section.

## Computational Details

All computations were performed in a development version of the Q-Chem software package.[73] The perfect pairing (PP) procedure starts with Pipek-Mezey localization of the Hartree-Fock orbitals, followed by Sano determination of initial virtual orbitals and full orbital optimization under the pairing ansatz.[74–81] Geometries for each of the molecules were optimized using the resolution-of-the-identity approach (RI) and the cc-pVTZ basis[82] combined with the RIMP2-cc-pVTZ auxiliary basis.[83]

iFCI computations were performed up to the $n = 4$ level. For each incremental term, a heat bath configuration interaction (HBCI) solver was used to compute the correlation energy $E_c(X)$. This method is discussed extensively in refs 33, 36 and 50-52. HBCI depends on convergence parameters called $\varepsilon$, which control HBCI's approach to the FCI limit. Herein, $\varepsilon_1$ was generally set to 0.5 mHa, and $\varepsilon_2$ to 0.1 μHa, which correspond to the variational and perturbative steps, respectively with deviations reported in the SI.

iFCI utilizes a convergence parameter ($\zeta$) which controls inclusion of virtual NOs in each incremental term (see ref 38). A full list of $\zeta$ values for all systems of interest is reported in the SI. The 3-body terms of the iFCI expansion were screened using the procedure described in reference 44. Based on the three 2-body terms that comprise a given 3-body term $E_c(ijk) \rightarrow \epsilon_{ij},\ \epsilon_{ik}$ and $\epsilon_{jk}$, at least two of the three must have magnitudes above a threshold $\mathcal{C}$ for the 3-body term to be significant. Otherwise, the 3-body term is excluded as low magnitude.

## **Results and Discussion**

### *Balancing Cost and Accuracy in iFCI and iNO-FCI*

To compare virtual orbital procedures for iFCI, the SNO method is compared with the proposed iNO strategy. Figure 2 compares the two procedures by showing the total energy as a function of $\zeta$ for cis-2-butene (24 valence electrons) in the cc-pVTZ (232 basis functions) basis set and n-octane (42 valence electrons) in the cc-pVTZ basis (380 basis functions). At the same $\zeta$ value, the cost of iNO-FCI is greater than that of traditional iFCI due to the need to compute $\mathrm{n}-1, \mathrm{n}-2\ldots$ terms whenever each n-body term is calculated. However, since iNO-FCI can converge more quickly with respect to $\zeta$, the iNO procedure can produce substantial cost savings in addition to increased accuracy. Figure 2 reports the least-expensive method that achieved convergence below chemical accuracy for each system.

The iNO procedure affords one additional benefit: each n-body term can be calculated using a different $\zeta$ threshold. This is possible because each $\epsilon_\mathrm{x}$ term is completely independent, c.f. equations 2-4, where (for example) $\epsilon_\mathrm{i}|\zeta_\mathrm{i}$ does not appear in equations 3 or 4. It is therefore possible to use tight $\zeta$ thresholds at low $n$-body level and then reduce $\zeta$ for higher $n$ when costs increase. This is the variable zeta method in Figure 2, where the 1- and 2-body correlation energies were found using $\zeta=10^{-10.5}$ , and the 3-body $\zeta$ is indicated in the legend. Convergence with respect to $\zeta$ occurs more quickly and the cost increase is marginal. For the two systems considered, using a variable $\zeta$ is the least computationally costly method to achieve chemical accuracy.

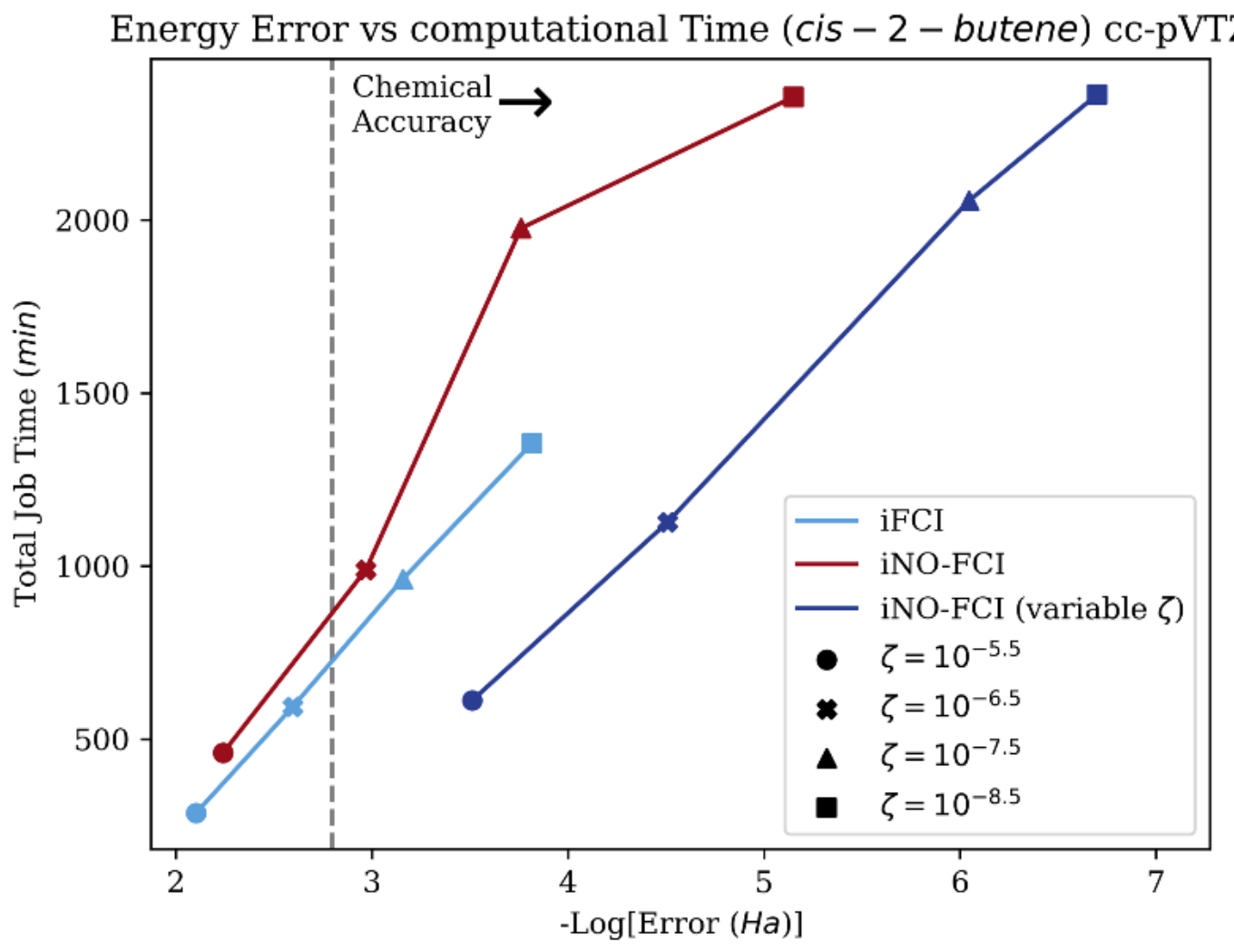

| | **cis-2-butene** | **n-octane** |
|---|---|---|
| **Best Method**[a] | iNO-FCI (variable ζ) | iNO-FCI (variable ζ) |
| **-log[ζ]** | 5.5 | 5.5 |
| **-log[error] (Ha)** | 3.5 | 3.3 |
| **Job Time (min)** | 612.2 | 6469.5 |
| **Valence e⁻** | 24 | 50 |
| **Basis Functions** | 232 | 380 |

[a] **See caption**

**Figure 2.** Total energy error compared vs computational time over a range of $\zeta$ values for cis-2-butene for iFCI, iNO-FCI, and iNO-FCI (variable $\zeta$) in the cc-pVTZ basis set. The table reports the least computationally expensive method that achieves chemical accuracy (1.6 mHa).

n-octane was also considered to confirm the trends found for cis-2-butene. In the cc-pVTZ basis, the computational cost of running iNO-FCI using a variable $\zeta=10^{-5.5}$ is lower than traditional iFCI with $\zeta=10^{-6.5}$. While the total energies from these results cannot be compared to a benchmark calculation (i.e. tight $\zeta$) due large system size and concomitant costs, they can be compared to one another. The difference between the two zeta values is $5 \cdot 10^{-4}$ Ha, below the threshold for chemical accuracy of 1 kcal/mol. Traditional iFCI has a difference of $9 \cdot 10^{-3}$ Ha, a factor nearly 6 times chemical accuracy. As such, a much higher value of $\zeta$ would be required to achieve chemical accuracy using traditional iFCI, consistent with the results for cis-2-butene. As the basis set increases in size, the cost savings of the iNO procedure is expected to be amplified, thus the cost-savings of the iNO procedure even more substantial.

## *Hydrocarbon scaling*

The computational cost of iFCI was empirically tested using a series of n-alkanes. The hydrocarbons under consideration were: $C_8H_{18}$, $C_{12}H_{26}$, $C_{16}H_{34}$, and $C_{20}H_{42}$, where the largest system contained 162 electrons. n-alkanes represent a best-case-scenario for testing scaling as the localized orbitals are straightforward to optimize and electronic correlation is expected to be nearly completely captured by the 3-body level. While many 3-body terms are expected to contribute a significant amount to the correlation energy, a great majority will involve groups of distant orbitals yielding minimal correlation energy. As such, 3-body screening depends on a single parameter $c$, to eliminate negligible 3-body terms (see computational details). Figure 3 shows that the screened 3-body iFCI requires approximately $N_e^{2.8}$ computational effort for iFCI ($n = 3$), down from an estimated $N_e^{4.4}$ for the unscreened computation. Our investigations indicated that to get convergence within chemical accuracy, the $\zeta$ for conventional iFCI must be ~$10^3$-fold tighter than that of iNO-FCI (see figure 2). As such, $\zeta = 10^{-5.5}$ was chosen for iNO-FCI and $\zeta = 10^{-8.5}$ was chosen for iFCI.

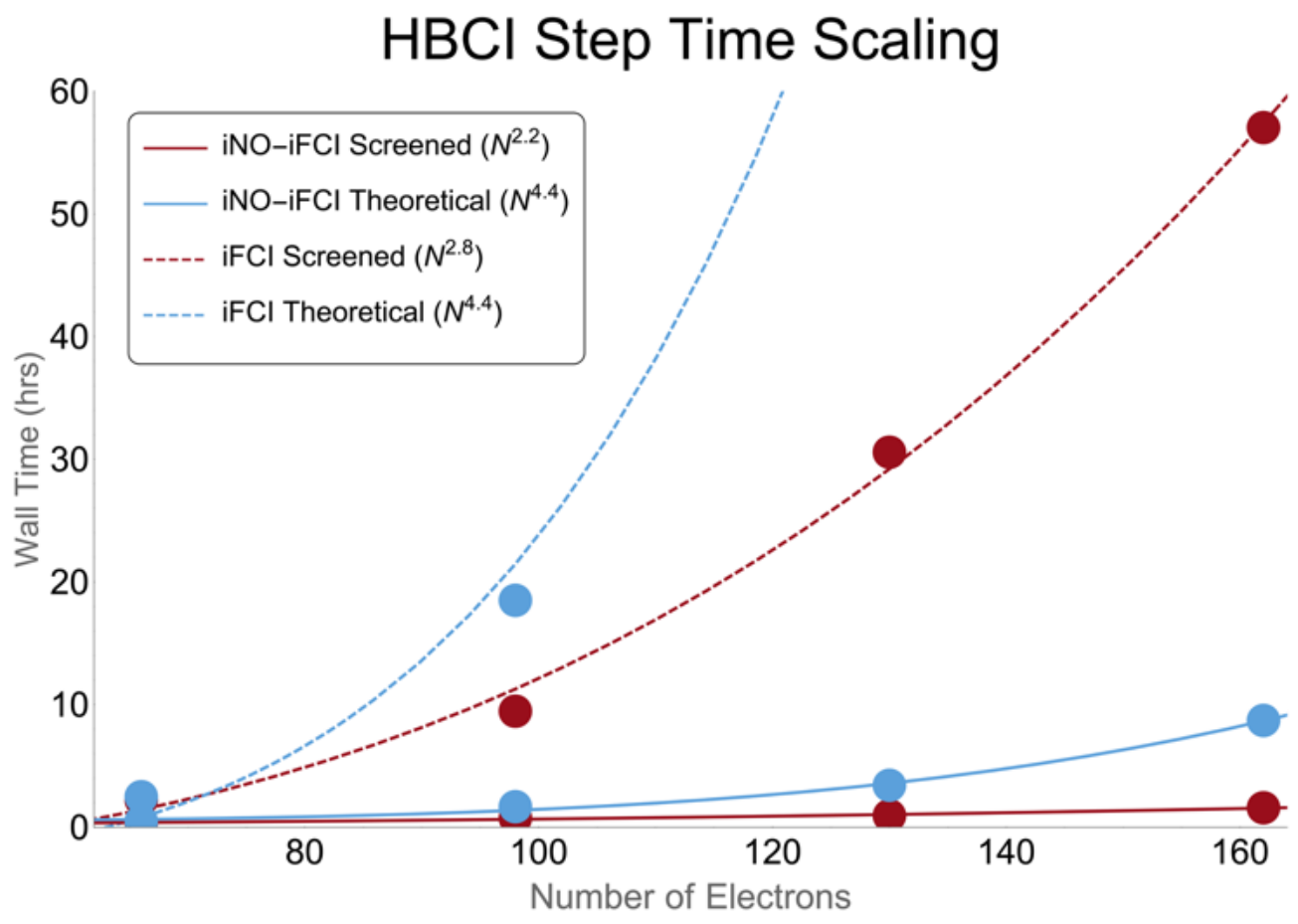

| System | n=3 energy (Ha) $\mathcal{C} = 10^{-6}$ | n=3 energy (Ha) $\mathcal{C} = 10^{-7}$ | Difference (kcal/mol) |
|---|---|---|---|
| $C_8H_{18}$ | -314.630 | -314.629 | 0.511 |
| $C_{12}H_{26}$, | -471.353 | -471.352 | 0.216 |
| $C_{16}H_{34}$ | -628.069 | -628.069 | 0.427 |
| $C_{20}H_{42}$ | -784.774 | -784.773 | 0.360 |

**Figure 3.** Total computational cost of the HBCI step of iFCI. The 3-body terms are screened (red) using the traditional iFCI (dotted) and iNO-FCI (solid), so the theoretical time for the unscreened calculation is also estimated (blue). The latter is computed as $t_{tot} = \left(\frac{N_{tot}}{N_{calc}}\right) t_{calc}$, where $t$ is the time, $N_{calc}$ is the number of terms with screening, and $N_{tot}$ is the full number of 3-body terms. Convergence was verified by reducing the screening parameter ($\mathcal{C}$) by a factor of 10, ensuring errors remained below chemical accuracy (1 kcal/mol).

Figure 3 indicates that not only does iNO-FCI scale more favorably than conventional iFCI, the prefactor is also significantly smaller resulting in a much less expensive calculation. The screening of 3-body terms greatly reduces the cost of iFCI with minimal impact on the overall result as shown in Figure 3. The $N^{2.2}$ scaling of iNO-FCI reflects the locality of electron correlation in hydrocarbons, as there is little correlation present beyond the 2-body level. Systems with longer-range many-body correlations are likely to show scaling factors higher than those shown here. In addition, the present implementation needs to compute electron repulsion integrals at cost $O(N^5)$, regardless of screening. At the present system sizes this cost is insignificant due to the use of the resolution-of-the-identity approximation.

*Bond Dissociation of $C_4H_6$ (cc-pVTZ)*

Energy profiles for bond dissociations are common test cases for strongly correlated wavefunction theories,[17,84] and iFCI has been successful in delineating several examples (see Ref 51). These calculations provide insight into reaction energetics, radical formation and the electronic structure in the bond-breaking process. High-bond-order dissociations are particularly challenging due to the emergence of multiradical character as bonds break, which single-reference electronic structure methods such as CCSD(T) typically fail to capture correctly.

To showcase the effectiveness of iNO-FCI on high-order bond dissociations, the dissociation of cis-2-butene along its central C=C bond was modeled in the cc-pVTZ and cc-pVQZ basis sets. This system represents an ideal test case due to its well-defined electronic structure at equilibrium and the significant strong correlation effects resulting from bond stretching. By comparing multiple truncation levels of iNO across different basis sets, we assess the method's ability to systematically recover correlation energy and accurately describe the dissociation curve.

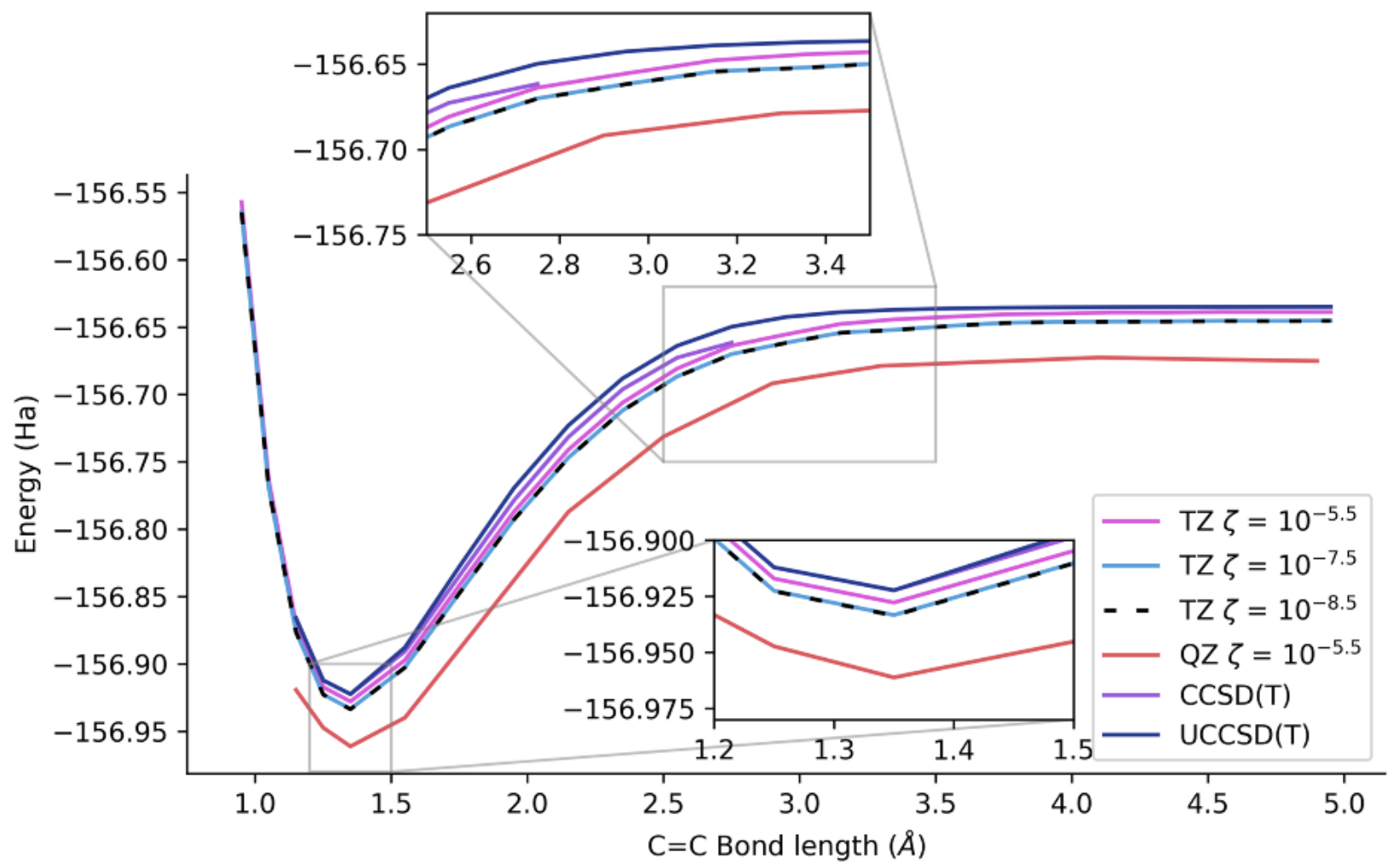

| Method[a] | Min Error (kcal/mol) | Max Error (kcal/mol) | Diff (kcal/mol) |
|---|---|---|---|
| **TZ $\zeta = 10^{-5.5}$** | 3.45 | 4.94 | 1.48 |
| **TZ $\zeta = 10^{-7.5}$** | 0.07 | 0.14 | 0.07 |
| **UCCSD(T)** | 6.52 | 15.10 | 8.58 |

[a] Compared to TZ $\zeta = 10^{-8.5}$

**Figure 4.** C=C bond dissociation of cis-2-butene using iNO-FCI at three values of ζ, compared to CCSD(T) (purple) and UCCSD(T) (dark blue) in the cc-pVTZ basis. Additionally, iNO-FCI in the cc-pVQZ basis with $\zeta = 10^{-5.5}$(red) is included. The table quantifies minimum and maximum deviations from the $\zeta = 10^{-8.5}$(dotted black) reference, showing that $\zeta = 10^{-7.5}$(light blue) and $\zeta = 10^{-5.5}$ (pink) primarily result in shifted curves due to basis set incompleteness, whereas UCCSD(T) exhibits significant deviations at ~2–3 times the equilibrium bond length.

In the TZ basis, the dissociation profile of iNO-FCI with $\zeta = 10^{-8.5}$ represents the most converged calculation and serves as the reference for defining the bond dissociation energy (BDE) of each method. While UCCSD(T) produces a BDE in reasonable agreement with the reference (Table 1), it exhibits significant deviations in the intermediate dissociation limit (c.f. Figure 4). Meanwhile, CCSD(T) fails to converge beyond approximately 2.8 Å. The iNO-FCI calculation with $\zeta = 10^{-7.5}$ agrees closely with the reference, indicating that the additional virtual functions included at $\zeta = 10^{-8.5}$ contribute minimally to the overall wavefunction. The $\zeta = 10^{-5.5}$ calculation exhibits a consistent shift in energy relative to $\zeta = 10^{-8.5}$, so the overall dissociation curve retains the same shape and a similar BDE is determined. This systematic behavior suggests that the iNO-FCI procedure enables reliable modeling of the dissociation in the cc-pVQZ basis with $\zeta = 10^{-5.5}$. Indeed, the BDE obtained in the QZ basis closely matches the results from the TZ basis, regardless of the ζ value used. Notably, even in the QZ basis, where cis-2-butene has 460 basis functions, the computational cost for $\zeta = 10^{-5.5}$ remains lower than for $\zeta = 10^{-7.5}$ or $\zeta = 10^{-8.5}$ in the TZ basis (Table 1).

The computational savings with using smaller $\zeta$ values are substantial. Each geometry required an average of 123.6 CPU hours for $\zeta = 10^{-8.5}$ but only 6.8 hours for $\zeta = 10^{-5.5}$, representing a 94% reduction in cost. Remarkably, iFCI with $\zeta = 10^{-5.5}$ is computationally less expensive than UCCSD(T) while achieving superior accuracy.

**Table 1.** BDE of iNO-FCI compared to that of UCCSD(T) and the average CPU time required per geometry.

| | BDE (kcal/mol) | CPU time (hr) |
|---|---|---|
| **TZ $\zeta = 10^{-5.5}$** | 181.38 | 6.8 |
| **TZ $\zeta = 10^{-7.5}$** | 180.81 | 104.4 |
| **TZ $\zeta = 10^{-8.5}$** | 180.81 | 123.6 |
| **QZ $\zeta = 10^{-5.5}$** | 179.53 | 64.8 |
| **TZ UCCSD(T)** | 180.37 | 18.2 |

*Singlet-Triplet Gaps of Highly Correlated Systems*

A prior study involving traditional iFCI (Ref 58) illustrated the ability of iFCI to accurately capture spin gaps for notorious high-correlation systems, where a polarized, TZ-quality basis was required to reach accurate energy gaps. This investigation was repeated here as a benchmark for improvements the iNO-FCI method can offer. The convergence of each of these chemical systems with respect to $\zeta$ was initially investigated. If the additional virtual functions changed the total energy of each spin state by less than 0.1 mHa, the relative spin gap was considered converged with respect to $\zeta$. Figure 6a illustrates how iNO improves convergence at low $\zeta$, where every system converged by $\zeta = 10^{-5.5}$. Alternatively, the original iFCI algorithm required an increasingly large ζ to achieve convergence as the size of the system increased. While a comprehensive investigation into convergence with respect to ζ was only done using the 6-31g* basis, we tested for the same trends with only a few select examples in the cc-pVTZ basis sets. As shown in the SI Section 4, these systems converged similarly to the smaller basis set.

iNO-FCI converges faster with respect to $\zeta$ and therefore reduces computational cost, but with what—if any—loss in accuracy? To address this question, a comparison of the agreement with experimental results of iFCI and iNO-FCI was warranted. For almost all systems shown in Figure 6, iNO-FCI does equally well or better compared to iFCI in its ability to accurately predict spin gaps.

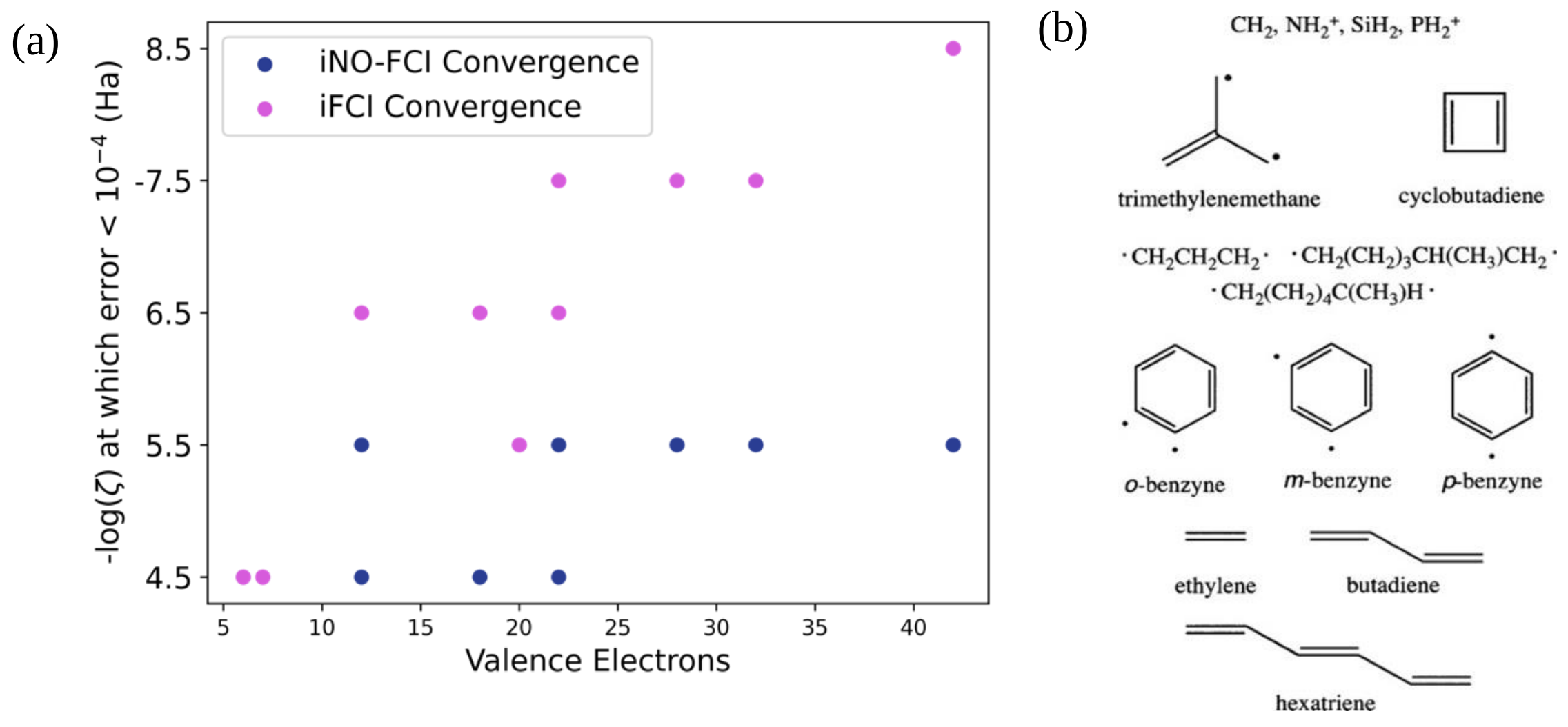

**Figure 5 a.)** The value of $\zeta$ where additional virtual functions yield less than 1 mHa as a function of system size. Results for 6-31g* basis for each system. **b.)** The benchmark systems under consideration.

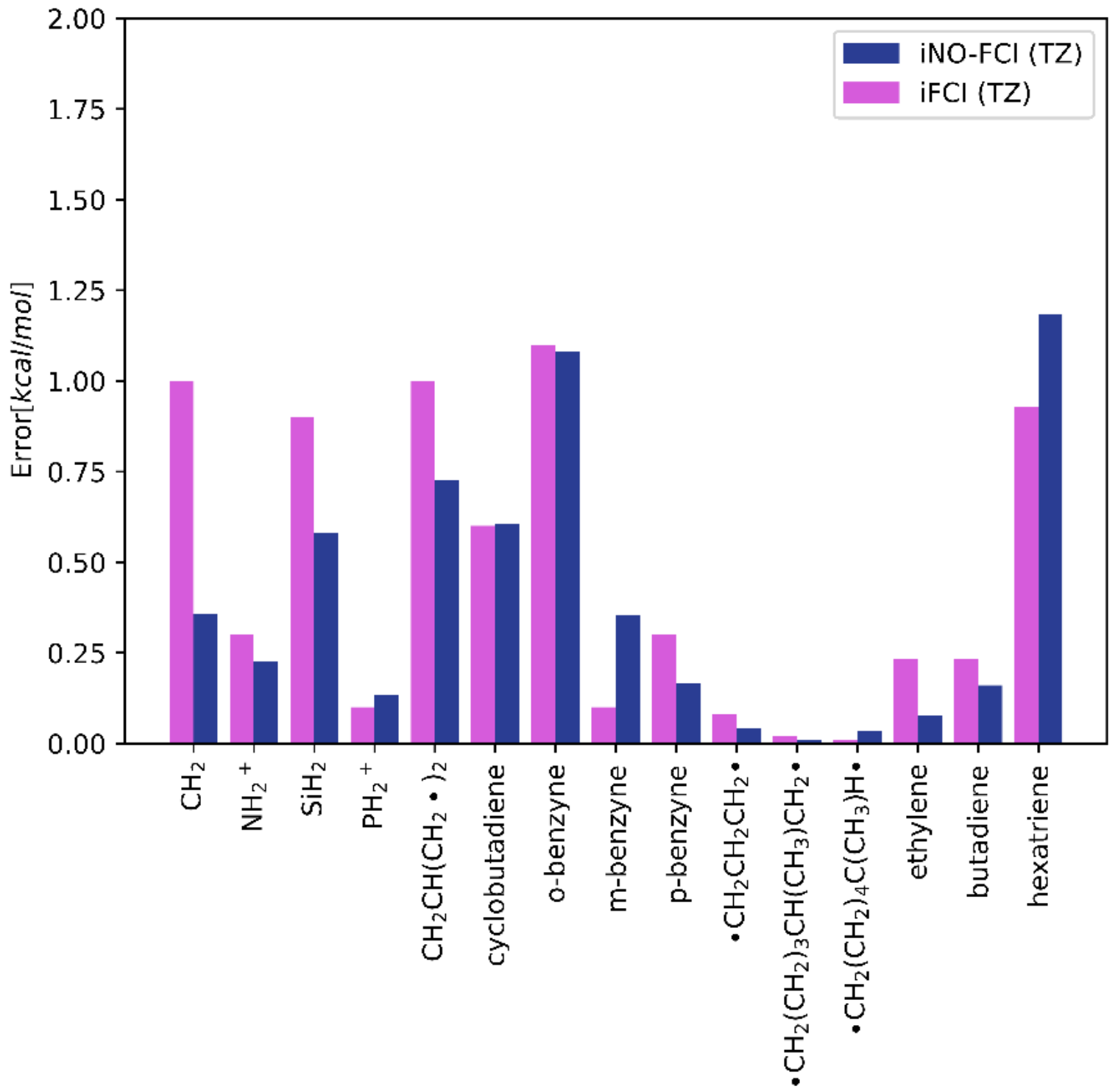

**Figure 6.** The error of iNO-FCI and iFCI in predicting the singlet triplet gap of the 15 biradical systems under consideration when compared to experimental values or the best electronic structure alternative. See Supplemental Information for details.

*Singlet Triplet Gap of Copper (II) Acetate Hydrate*

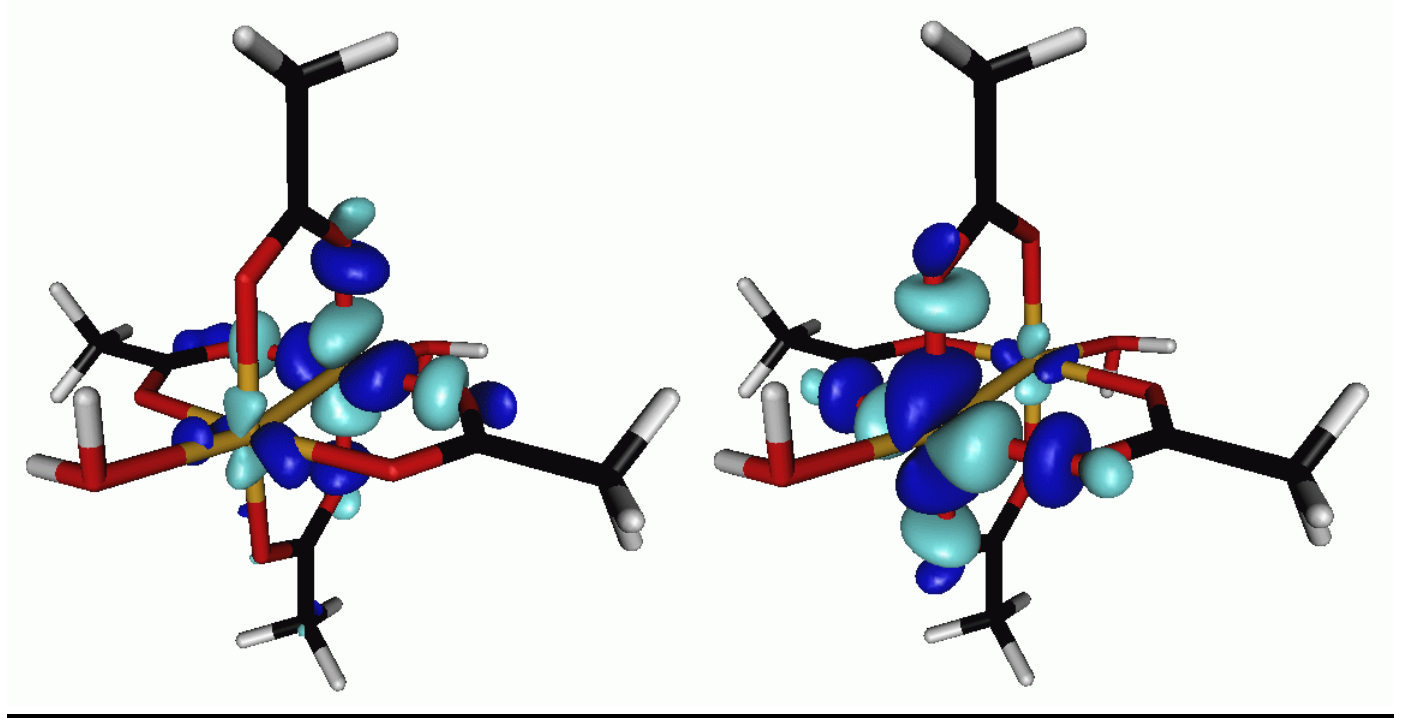

| iFCI Orbital Occupancies | | |
|---|---|---|
| | Singlet | Triplet |
| Orbital 101 (left) | 1.11 | 1.10 |
| Orbital 102 (right) | 1.05 | 1.05 |

**Figure 7.** The two singly occupied molecular orbitals from the 3-body natural orbital calculation responsible for the singlet-triplet gap in Cu(aqac) with corresponding eigenvalues representing orbital occupations.

Transition metal complexes are especially difficult to accurately model, as they require rigorous treatment of electronic correlation.[85–89] The copper(II) acetate hydrate (Cu(aqac)) complex is typically used as a model for measuring magnetic and electronic interactions between the copper centers.[85,90–99] Cu(aqac) is a biradical in the singlet state, arising from an unpaired d electron on each Cu atom (see Figure 7). The singlet and triplet states are therefore nearly degenerate, with a difference in energy of only 286 $cm^{-1}$.[100] Herein, iNO-FCI was tested against experiment and the prior iFCI method to quantify this spin gap. Since our previous calculation[56] was performed in the 6-31g* basis, we tested two values of $\zeta$ in the 6-31g* basis (all atoms) to compare convergence. We subsequently tested the resolution of the spin states using split cc-pVTZ(Cu)/6-31G*(all other) basis using $\zeta = 10^{-5.5}$.

**Table 2.** The singlet-triplet gap of copper(II) acetate hydrate using iFCI and iNO-FCI in various basis sets compared to other electronic structure methods and experiment.

| **Method** | **Basis** | **S-T gap ($cm^{-1}$)** |
|---|---|---|
| **iFCI ($\zeta = 10^{-5.5}$)** | 6-31g* | 117.8 |
| **iNO-FCI ($\zeta = 10^{-5.5}$)** | 6-31g* | 150.3 |
| **iNO-FCI ($\zeta = 10^{-6.5}$)** | 6-31g* | 222.6 |
| **iNO-FCI ($\zeta = 10^{-5.5}$)** | cc-pVTZ(Cu)/6-31G*(other) | 289.0 |
| **Experiment[a]** | - | 286 |
| **EOM-SF-CCSD[b]** | cc-pVTZ | 180 |
| **UHF EOM-SF-CCSD[c]** | cc-pVTZ | 191 |
| **AP-UCCSD[d]** | 6-31g | 190 |

a) Reference(99)
b) Reference(89)
c) Reference(91)

d) Reference(97)

This calculation represented a significant computational undertaking. The large number of localized molecular orbitals centered on the metal atoms and surrounding ligands are highly correlated. In the 6-31G* basis, the 374 basis functions and 130 valence electrons generate approximately $10^{141}$ electronic configurations in FCI. Expanding to the hybrid basis with 420 basis functions increases the CI dimension to $10^{147}$ determinants.

As shown in Table 2, iNO-FCI in the 6-31G* basis provides a better resolution of the gap compared to the traditional iFCI algorithm. Using the split cc-pVTZ(Cu)/6-31G*(all other) basis, the gap is further refined, and the error relative to experimental values drops below 10 $cm^{-1}$. This improvement is likely due to the increased number of polarization functions included for each copper atom in the larger basis, which better captures correlation effects that the 6-31G* basis struggles to resolve.

*Reaction Mechanism of Criegee Intermediate*

Criegee intermediates are a class of zwitterionic, biradical molecules with significant importance to nighttime oxidation reactions in the atmosphere.[101,102] The most fundamental Criegee intermediate ($CH_2OO$) represents an excellent test for evaluating electronic structure methods. The bond stretching and electron density delocalization inherent in transition states exacerbate the high degree of static correlation in reactions involving Criegee intermediates. Obtaining accurate rate constants for atmospheric chemistry models depends sensitively on the activation energy barrier, so even small deficiencies in the method can be problematic. As a prototypical example of a Criegee reaction, $CH_2OO+H_2O \rightarrow HOCH_2OOH$ (see Figure 8) is investigated.[63,103] A prior study of this reaction using QCISD qualitatively captured the mechanism but differed from the iFCI results quantitatively. iNO-FCI was employed up to the 3-body level (no screening) using $\zeta = 10^{-6.5}$ in the cc-pVDZ and cc-pVTZ basis sets (see additional details in SI). The QCISD path differs from iFCI/TZ by at least 3 kcal/mol in the relative energies of intermediate and transition states, and by nearly 10 kcal/mol in the final product energy. UCCSD(T) agrees more closely with iFCI/TZ in that the transition state and intermediate energies differ by approximately 0.5 kcal/mol and the final product energy differs by 2.5 kcal/mol. Depending on the functional employed, an energy range of between 6 kcal/mol (for the intermediate) to above 25 kcal/mol (for the product) is found. This highlights the need for a high-accuracy method, such as iFCI, to capture the strong correlation effects involved in this reaction.

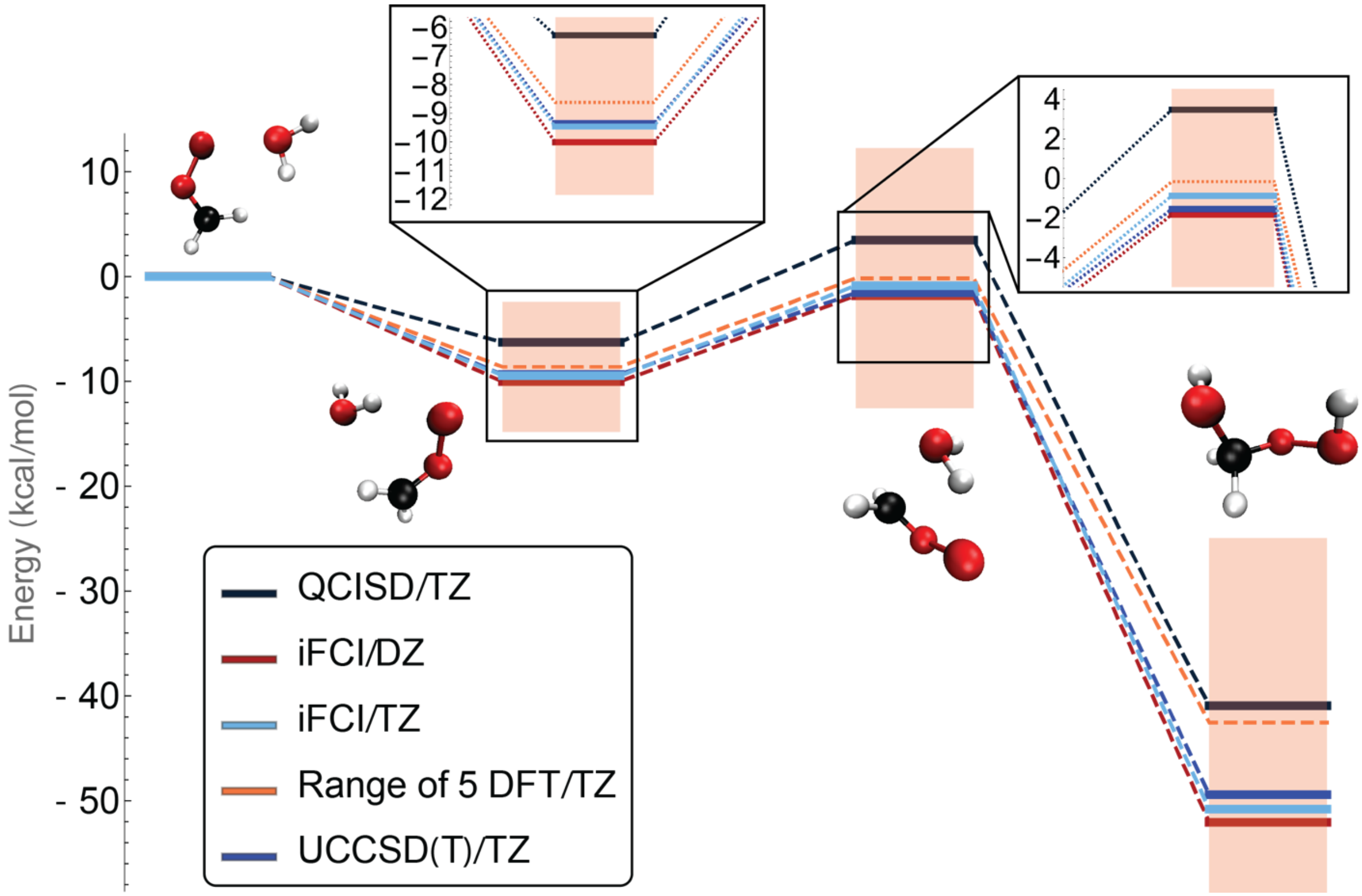

**Figure 8.** iNO-FCI applied to a reaction involving the criegee intermediate with water demonstrating the necessity of modeling highly correlated systems with high accuracy methods. The range results from 5 DFT functionals (PBE, B97-D, M06-2X, B3LYP, and WB97X-D each with stability analysis), UCCSD(T), and QCISD included for reference.

## Conclusion

Traditional electronic structure algorithms struggle with the high computational costs of accurately modeling strongly correlated systems, especially when basis sets beyond double-zeta are employed. Herein, iNO-FCI was shown capable of handling weakly and strongly correlated molecules in polarized triple- and quadruple-zeta basis sets. The polynomial cost of iNO-iFCI allowed investigation of a transition metal complex with over 100 electrons, with better convergence than the prior iFCI approach. In all, the method is effective in modeling potential energy surfaces, bond dissociation profiles, and spin gaps of characteristically complex, multi-reference systems. These advancements expand the range of systems that can be studied with near-FCI accuracy.

## Supporting Information

The supporting information contains:

- iFCI energies and CPU timings for each value of $\zeta$ for cis-2-butene and n-octane.
- Data used to generate Figure 3 of the iFCI calculation of $C_8H_{18}$, $C_{12}H_{26}$, $C_{16}H_{34}$, and $C_{20}H_{42}$.
- Energies for the bond dissociation of cis-2-butene using iFCI and CCSD(T).

- Singlet and triplet state energy values for each value of $\zeta$ for the 15 highly correlated systems
- Singlet and triplet state iFCI energy values for copper (II) acetate hydrate.
- The energies of the systems involved in the Criegee intermediate-water reaction in iFCI, 5 DFT functionals (PBE, B97-D, M06-2X, B3LYP, and $\omega$B97X-D each with stability analysis) and UCCSD(T)

**Acknowledgments**
The authors would like to thank the U.S. Department of Energy, Office of Science, Basic Energy Sciences for their support of this work through DE-SC0022241. Our thanks also to David Braun who continuously supports our computational endeavors.

TOC Graphic

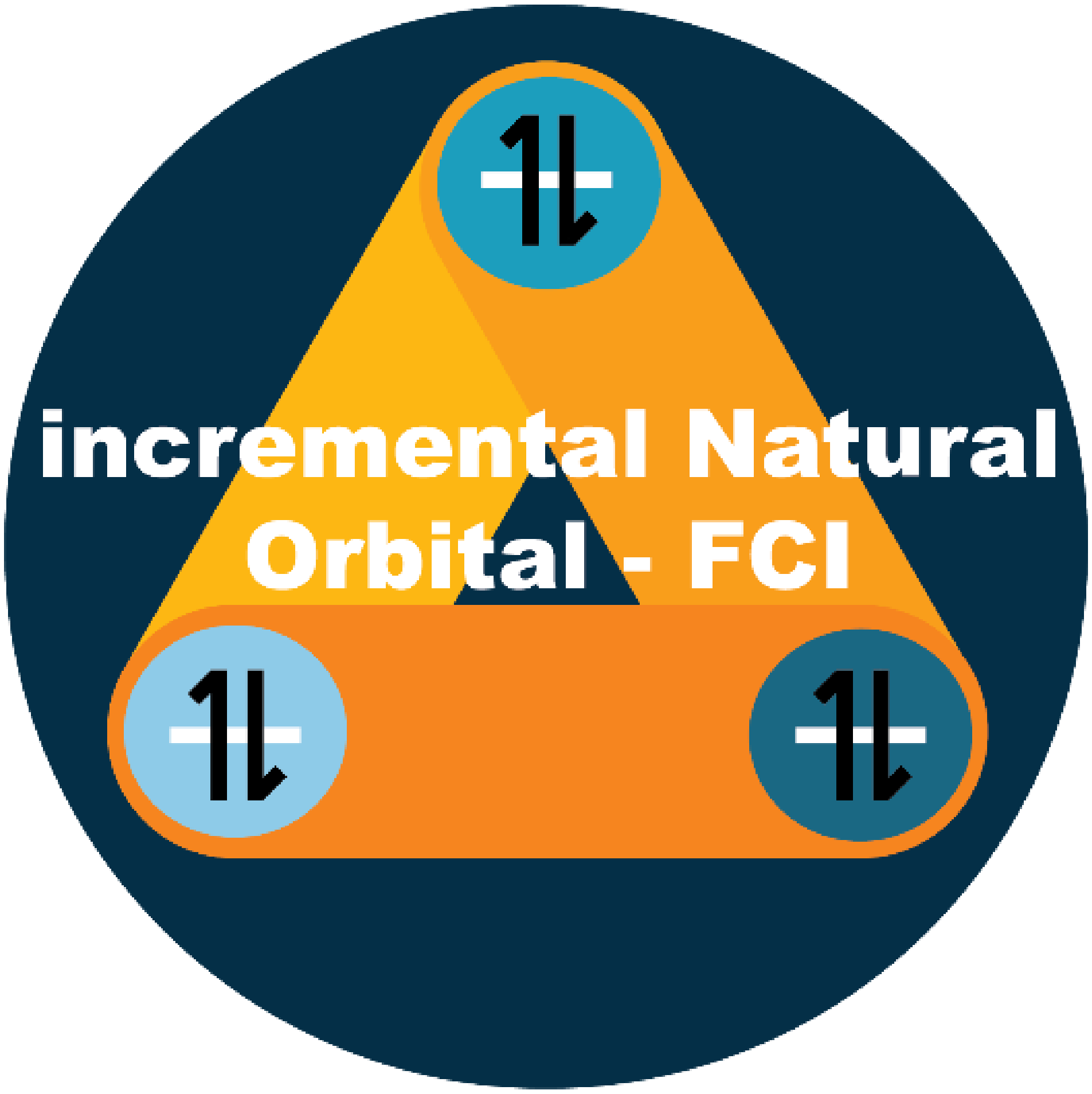

Supporting Information for
Taming the Virtual Space for Incremental Full Configuration Interaction

Jeffrey Hatch, Paul M. Zimmerman*
Department of Chemistry, University of Michigan
930 N. University Ave, Ann Arbor, MI 48109, USA, *paulzim@umich.edu

Table of Contents:

## I. Balancing Cost and Accuracy in iFCI and iNO-FCI

The timing data used to determine the convergence with respect to $\zeta$ for cis-2-butene and n-octane are represented in Table 1.

**Table 1.** The convergence of energy with respect to ζ for n-octane and cis-2-butene with accompanying computational times. Basis sets shorthand - DZ: cc-pVDZ, TZ: cc-pVTZ.

| Compound | Basis | Method | -log(ζ) | n = 2 (Ha) | n = 3 (Ha) | n = 2 time (min) | n = 3 time (min) | Total CPU Time (min) |
|---|---|---|---|---|---|---|---|---|
| **cis-2-butene** | DZ | iFCI | 5.5 | -156.7292 | -156.7636 | 0.68 | 15.98 | 17.06 |
| **cis-2-butene** | DZ | iFCI | 6.5 | -156.7311 | -156.7661 | 1.20 | 37.70 | 39.31 |
| **cis-2-butene** | DZ | iFCI | 7.5 | -156.7314 | -156.7664 | 1.98 | 47.00 | 49.42 |
| **cis-2-butene** | DZ | iFCI | 8.5 | -156.7314 | -156.7664 | 2.39 | 49.03 | 51.90 |
| **cis-2-butene** | DZ | iFCI | 9.5 | -156.7314 | -156.7664 | 2.39 | 49.03 | 51.89 |
| **cis-2-butene** | DZ | iNO-FCI | 5.5 | -156.7286 | -156.7677 | 0.48 | 10.34 | 11.23 |
| **cis-2-butene** | DZ | iNO-FCI | 6.5 | -156.7300 | -156.7639 | 0.93 | 23.98 | 25.33 |
| **cis-2-butene** | DZ | iNO-FCI | 7.5 | -156.7302 | -156.7622 | 1.55 | 29.99 | 31.99 |
| **cis-2-butene** | DZ | iNO-FCI | 8.5 | -156.7302 | -156.7620 | 1.86 | 30.90 | 33.23 |
| **cis-2-butene** | DZ | iNO-FCI | 9.5 | -156.7302 | -156.7621 | 1.88 | 30.92 | 33.28 |
| **cis-2-butene** | TZ | iFCI | 5.5 | -156.8871 | -156.9278 | 13.96 | 445.42 | 477.54 |
| **cis-2-butene** | TZ | iFCI | 6.5 | -156.8915 | -156.9325 | 28.29 | 959.85 | 1007.76 |
| **cis-2-butene** | TZ | iFCI | 7.5 | -156.8924 | -156.9334 | 85.18 | 1888.87 | 1993.99 |
| **cis-2-butene** | TZ | iFCI | 8.5 | -156.8925 | -156.9336 | 152.08 | 2195.66 | 2372.97 |
| **cis-2-butene** | TZ | iFCI | 9.5 | -156.8926 | -156.9336 | 152.00 | 2190.36 | 2372.79 |
| **cis-2-butene** | TZ | iNO-FCI | 5.5 | -156.8885 | -156.9390 | 9.85 | 275.26 | 303.43 |
| **cis-2-butene** | TZ | iNO-FCI | 6.5 | -156.8914 | -156.9336 | 22.37 | 569.81 | 611.55 |
| **cis-2-butene** | TZ | iNO-FCI | 7.5 | -156.8919 | -156.9318 | 56.26 | 904.15 | 978.18 |
| **cis-2-butene** | TZ | iNO-FCI | 8.5 | -156.8920 | -156.9309 | 110.93 | 1235.24 | 1373.15 |
| **cis-2-butene** | TZ | iNO-FCI | 9.5 | -156.8920 | -156.9311 | 110.11 | 1230.21 | 1369.86 |
| **n-octane** | DZ | iFCI | 5.5 | -309.3977 | -309.8402 | 0.04 | 5590.48 | 5590.93 |
| **n-octane** | DZ | iFCI | 6.5 | -309.3977 | -309.8407 | 0.04 | 8144.99 | 8145.44 |
| **n-octane** | DZ | iFCI | 7.5 | -309.3977 | -309.8408 | 0.00 | 8249.88 | 8250.41 |
| **n-octane** | DZ | iFCI | 8.5 | -309.3977 | -309.8408 | 0.00 | 8229.08 | 8229.56 |
| **n-octane** | DZ | iFCI | 9.5 | -309.3977 | -309.8408 | 0.00 | 8202.51 | 8202.98 |
| **n-octane** | DZ | iNO-FCI | 5.5 | -309.4035 | -309.9709 | 96.57 | 3475.38 | 3597.90 |
| **n-octane** | DZ | iNO-FCI | 6.5 | -309.4001 | -309.8387 | 190.41 | 5082.47 | 5300.81 |
| **n-octane** | DZ | iNO-FCI | 7.5 | -309.3968 | -309.8627 | 208.84 | 5063.61 | 5302.37 |
| **n-octane** | DZ | iNO-FCI | 8.5 | -309.3967 | -309.8641 | 208.22 | 5056.43 | 5295.74 |
| **n-octane** | DZ | iNO-FCI | 9.5 | -309.3966 | -309.8642 | 208.81 | 5122.14 | 5362.83 |
| **n-octane** | TZ | iFCI | 5.5 | -310.1235 | -310.2124 | 158.23 | 6153.49 | 6469.54 |
| **n-octane** | TZ | iFCI | 6.5 | -310.1364 | -310.2258 | 136.55 | 8476.26 | 8700.21 |
| **n-octane** | TZ | iNO-FCI | 5.5 | -310.1327 | -310.2443 | 43.90 | 1851.81 | 1975.71 |
| **n-octane** | TZ | iNO-FCI | 6.5 | -310.1387 | -310.2352 | 110.26 | 5484.14 | 5682.58 |

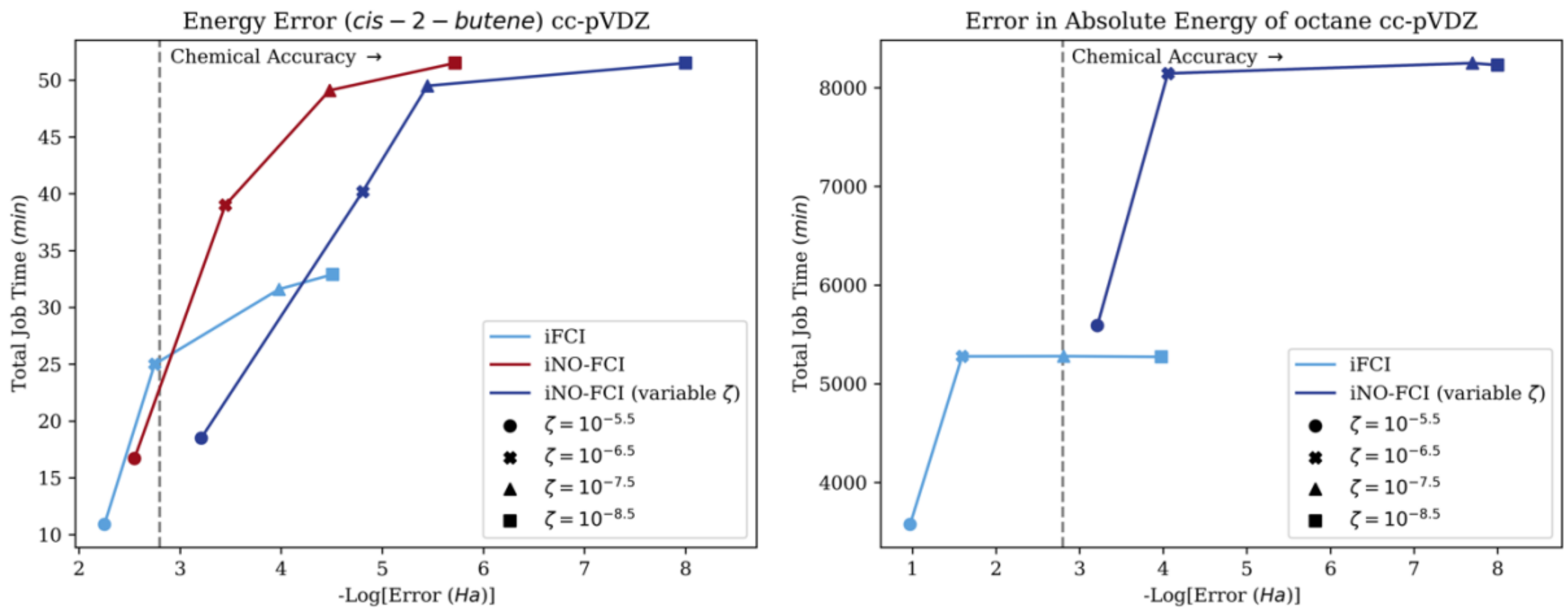

**Figure 1.** Error compared to near-complete virtual space calculation vs time with respect to each $\zeta$ value for cis-2-butene(left) and octane (left). iFCI, iNO-FCI and iNO-FCI (variable $\zeta$) are compared in the DZ basis set. Note: only the iNO-FCI (variable $\zeta$) was utilized for n-octane.

The pattern for error vs. $\zeta$ in n-octane in the DZ basis is different than that of cis-2-butene. As expected, the iNO procedure gives a more converged result than traditional iFCI for each value of $\zeta$ and the computational cost is also greater. However, unlike in the TZ basis, the least computationally expensive calculation in the DZ basis for n-octane that is below chemical accuracy is $\zeta = 10^{-8.5}$ using traditional iFCI. This differs from the results of the TZ basis reported in the main body of this work where the least computationally expensive method to reach chemical accuracy was iNO-FCI (variable zeta) where $\zeta = 10^{-5.5}$. As mentioned in the main body of this work, the computational cost savings of the iNO-FCI method increases with basis size. For n-octane, the cc-pVDZ basis is insufficiently large to benefit from the iNO method compared to the traditional iFCI algorithm. The smaller system cis-2-butene follows the same pattern in both basis sets. He authors hypothesize that the low levels of correlation in n-alkanes might be the reason for this discrepancy.

## II. Hydrocarbon scaling

The hydrocarbons under consideration for timings of iFCI and iNO-FCI were*:* $C_8H_{18}$, $C_{12}H_{26}$, $C_{16}H_{34}$, and $C_{20}H_{42}$. The computational time associated with each step in the process (generating integrals, computing the Fock matrix, and the HBCI time) were considered. While this study pointed to areas for improvement in the integral transforms and Fock build steps, the authors did not feel reporting these in the main manuscript was warranted as the iNO procedure does not improve upon these steps. Future work will seek to improve the overall scaling of the method by optimizing the code that computes molecular orbital integrals and generates Fock matrices. All geometries were optimized using B3LYP/6-31g* level of theory.

**Table 2**. The CPU time to compute each step in the 3-body terms as well as screening information for each of the 4 alkanes considered, $C_8H_{18}$, $C_{12}H_{26}$, $C_{16}H_{34}$, and $C_{20}H_{42}$.

| | **iFCI** | **iNO-FCI** |
|---|---|---|

| | $C_8H_{18}$ | $C_{12}H_{26}$ | $C_{16}H_{34}$ | $C_{20}H_{42}$ | $C_8H_{18}$ | $C_{12}H_{26}$ | $C_{16}H_{34}$ | $C_{20}H_{42}$ |
|---|---|---|---|---|---|---|---|---|
| **Valence Electrons** | 50 | 74 | 98 | 122 | 50 | 74 | 98 | 122 |
| **3-body terms calculated** | 2056 | 3974 | 6575 | 8961 | 1658 | 3497 | 5104 | 6561 |
| **Possible 3-body terms** | 2300 | 7770 | 18424 | 35990 | 2300 | 7770 | 18424 | 35990 |
| **Integral time (hr)** | 1.9 | 19.7 | 114.8 | 437.5 | 0.3 | 1.2 | 2.4 | 5.6 |
| **Fock time (hr)** | 0.2 | 0.6 | 1.7 | 3.6 | 0.9 | 3.9 | 7.8 | 18.6 |
| **HBCI time (hr)** | 2.2 | 9.5 | 30.6 | 57.0 | 0.3 | 0.8 | 0.9 | 1.6 |

## III. Bond Dissociation of $C_4H_6$ (cc-pVTZ)

Perfect Pairing (PP) orbitals at the equilibrium bond length of cis-2-butene were generated and used as a starting point for the the geometries. As the bond was stretched or contracted, the orbitals from the previous geometry were read in and then optimized at the current geometry. In the iFCI and CCSD(T) calculations in the cc-pVTZ (TZ in the table) basis, passing the orbitals from the previous geometry every 0.2 Å was sufficient to achieve a smooth dissociation curve. However, 0.2 Å yielded an abnormally high BDE in the cc-pVQZ (QZ in the table below) basis. When orbitals were optimized every 0.05 Å, the BDE achieved the expected value, suggesting that the orbitals were not fully continuous when generated in the larger step size. Due to the larger computational costs of the iFCI calculation in the QZ basis, the subset of QZ geometries was smaller than that of the TZ basis.

**Table 3**. The energy values for the bond dissociation of cis-2-butene using iFCI in the cc-pVTZ and cc-pVQZ basis sets with different values of $\zeta$.

| Bond Length | iFCI TZ $\zeta=10^{-5.5}$ | iFCI TZ $\zeta=10^{-7.5}$ | iFCI TZ $\zeta=10^{-8.5}$ | iFCI QZ $\zeta=10^{-5.5}$ | TZ CCSD(T) | TZ UCCSD(T) |
|---|---|---|---|---|---|---|
| **0.95** | -156.5572 | -156.5645 | -156.5647 | - | -156.8659 | - |
| **1.05** | -156.7635 | -156.7696 | -156.7698 | - | -156.9121 | - |
| **1.15** | -156.8700 | -156.8761 | -156.8763 | -156.9192 | -156.9223 | -156.8659 |
| **1.25** | -156.9171 | -156.9225 | -156.9227 | -156.9473 | -156.8906 | -156.9121 |
| **1.357 (equilibrium)** | -156.9278 | -156.9333 | -156.9336 | -156.9611 | -156.8345 | -156.9223 |
| **1.55** | -156.8971 | -156.9024 | -156.9026 | -156.9390 | -156.7787 | -156.8877 |
| **1.75** | -156.8416 | -156.8473 | -156.8475 | - | -156.7316 | -156.8276 |
| **1.95** | -156.7870 | -156.7928 | -156.7929 | - | -156.6960 | -156.7695 |
| **2.15** | -156.7411 | -156.7469 | -156.7470 | -156.7871 | -156.6726 | -156.7230 |
| **2.35** | -156.7059 | -156.7117 | -156.7118 | - | -156.6615 | -156.6800 |
| **2.5** | - | - | - | -156.7312 | - | - |
| **2.55** | -156.6807 | -156.6860 | -156.6867 | - | - | -156.6638 |
| **2.75** | -156.6637 | -156.6700 | -156.6701 | - | - | -156.6497 |
| **2.9** | - | - | - | -156.6916 | - | - |
| **2.95** | -156.6550 | -156.6615 | -156.6617 | - | - | -156.6424 |
| **3.15** | -156.6476 | -156.6541 | -156.6542 | - | - | -156.6389 |
| **3.3** | - | - | - | -156.6787 | - | - |
| **3.35** | -156.6442 | -156.6519 | -156.6520 | - | - | -156.6370 |
| **3.55** | -156.6423 | -156.6490 | -156.6492 | - | - | -156.6361 |
| **3.75** | -156.6405 | -156.6468 | -156.6469 | - | - | -156.6356 |
| **3.95** | -156.6000 | -156.6460 | -156.6461 | - | - | -156.6353 |
| **4.1** | - | - | - | -156.6725 | - | - |
| **4.15** | -156.6390 | -156.6458 | -156.6460 | - | - | -156.6351 |
| **4.35** | -156.6391 | -156.6450 | -156.6458 | - | - | -156.6350 |
| **4.5** | - | - | - | -156.6739 | - | - |
| **4.55** | -156.6386 | -156.6453 | -156.6450 | - | - | -156.6349 |
| **4.75** | -156.6387 | -156.6450 | -156.6455 | - | - | -156.6349 |
| **4.9** | - | - | - | -156.6750 | - | - |
| **4.95** | -156.6387 | -156.6452 | -156.6453 | - | - | -156.6348 |

## IV. Singlet-Triplet Gaps of Highly Correlated Systems

The singlet-triplet gaps of the 15 systems considered in the main manuscript are reported below.[1] In each case, the iNO-FCI was calculated with $-\log(\zeta) = 5.5, \epsilon_1 = 10^{-4}$ and $\epsilon_2 = 10^{-7}$. The details for the previous study calculating the same gaps using the traditional iFCI algorithm can be found in ref 1. For investigating convergence with respect to $\zeta$, the energies for these systems were recomputed with iFCI and iNO-FCI. Once the energy difference between two sequential values of $\zeta$ was below $10^{-4}$, the system was considered converged with respect to $\zeta$. Any gaps in Table 3 were a result in iFCI calculations that did not converge, but did not affect the determination of convergence outlined above and were therefore not included herein. Geometries for each system can be found in reference 1.

**Table 4.** The energies of the 15 systems under consideration at each value of $\zeta$ in the singlet and triplet spin states using the iNO-FCI and traditional iFCI algorithms in the 6-31g* basis. All energies in Ha. S refers to the singlet state and T refers to the triplet.

| | $\boldsymbol{\zeta =}$ | $\mathbf{10^{-4.5}}$ | $\mathbf{10^{-5.5}}$ | $\mathbf{10^{-6.5}}$ | $\mathbf{10^{-7.5}}$ | $\mathbf{10^{-8.5}}$ | $\mathbf{10^{-9.5}}$ |
|---|---|---|---|---|---|---|---|
| **iNO-FCI T** | $CH_2$ | -39.0751 | -39.0782 | -39.0785 | -39.0785 | -39.0786 | -39.0786 |
| **iNO-FCI T** | $NH_2^+$ | -55.3442 | -55.3442 | -55.3442 | -55.3442 | -55.3442 | -55.3442 |
| **iNO-FCI T** | $PH_2^+$ | -341.6344 | -341.6344 | -341.6344 | -341.6344 | -341.6344 | -341.6344 |
| **iNO-FCI T** | $SiH_2$ | -290.1082 | -290.1082 | -290.1082 | -290.1082 | -290.1082 | -290.1082 |
| **iNO-FCI T** | $C_2H_4$ | -78.1869 | -78.1856 | -78.1858 | -78.1859 | -78.1859 | -78.1859 |
| **iNO-FCI T** | $C_4H_6$ | -155.4115 | -155.4141 | -155.4165 | -155.4150 | -155.4173 | -155.4173 |
| **iNO-FCI T** | Propane | -117.4517 | -117.4521 | -117.4522 | -117.4522 | -117.4522 | -117.4522 |
| **iNO-FCI T** | TMM (triplet) | -155.4783 | -155.4792 | -155.4792 | -155.4777 | -155.4792 | -155.4792 |
| **iNO-FCI T** | Cyclobutadiene | -154.2374 | -154.2441 | -154.2443 | -154.2466 | -154.2392 | -154.2465 |
| **iNO-FCI T** | n-heptane | -274.2502 | -274.2519 | -274.2522 | -274.2524 | -274.2524 | -274.2524 |
| **iNO-FCI T** | 2-methyl-hexane | -274.1129 | -274.1143 | -274.1146 | -274.1147 | -274.1147 | -274.1147 |
| **iNO-FCI T** | m-benzyne | -230.2177 | -230.2192 | -230.2194 | -230.2195 | -230.2195 | -230.2195 |
| **iNO-FCI T** | o-benzyne | -230.2120 | -230.2135 | -230.2138 | -230.2139 | -230.2139 | -230.2139 |
| **iNO-FCI T** | p-benzyne | -230.2240 | -230.2255 | -230.2258 | -230.2258 | -230.2258 | -230.2258 |
| **iNO-FCI T** | $C_6H_{10}$ | -232.7223 | -232.6306 | -232.6309 | -232.6310 | -232.6310 | -232.6310 |
| **iNO-FCI S** | $CH_2$ | -39.0597 | -39.0623 | -39.0626 | -39.0627 | -39.0627 | -39.0627 |
| **iNO-FCI S** | $NH_2^+$ | -55.2946 | -55.2946 | -55.2946 | -55.2946 | -55.2946 | -55.2946 |
| **iNO-FCI S** | $PH_2^+$ | -341.6589 | -341.6589 | -341.6589 | -341.6589 | -341.6589 | -341.6589 |
| **iNO-FCI S** | $SiH_2$ | -290.1382 | -290.1382 | -290.1382 | -290.1382 | -290.1382 | -290.1382 |
| **iNO-FCI S** | $C_2H_4$ | -78.3575 | -78.3576 | -78.3576 | -78.3576 | -78.3576 | -78.3576 |
| **iNO-FCI S** | $C_4H_6$ | -155.5423 | -155.5427 | -155.5428 | -155.5428 | -155.5428 | -155.5428 |
| **iNO-FCI S** | Propane | -117.4504 | -117.4508 | -117.4509 | -117.4509 | -117.4509 | -117.4509 |
| **iNO-FCI S** | TMM (singlet) | -155.4532 | -155.4536 | -155.4538 | -155.4538 | -155.4538 | -155.4538 |
| **iNO-FCI S** | Cyclobutadiene | -154.3104 | -154.3110 | -154.3110 | -154.3110 | -154.3110 | -154.3110 |
| **iNO-FCI S** | n-heptane | -274.2529 | -274.2544 | -274.2547 | -274.2548 | -274.2549 | -274.2549 |
| **iNO-FCI S** | 2-methyl-hexane | -274.2093 | -274.2105 | -274.2108 | -274.2109 | -274.2109 | -274.2109 |
| **iNO-FCI S** | m-benzyne | -230.2499 | -230.2511 | -230.2514 | -230.2514 | -230.2514 | -230.2514 |
| **iNO-FCI S** | o-benzyne | -230.2660 | -230.2672 | -230.2675 | -230.2676 | -230.2676 | -230.2676 |
| **iNO-FCI S** | p-benzyne | -230.2320 | -230.2335 | -230.2338 | -230.2338 | -230.2338 | -230.2338 |
| **iNO-FCI S** | $C_6H_{10}$ | -232.7351 | -232.7362 | -232.7365 | -232.7365 | -232.7365 | -232.7365 |
| | $\zeta =$ | $10^{-4.5}$ | $10^{-5.5}$ | $10^{-6.5}$ | $10^{-7.5}$ | $10^{-8.5}$ | $10^{-9.5}$ |
| **iFCI T** | $CH_2$ | -39.0777 | -39.0785 | -39.0785 | -39.0785 | -39.0785 | -39.0785 |
| **iFCI T** | $NH_2^+$ | -55.3849 | -55.3892 | -55.3896 | -55.3897 | -55.3898 | -55.3898 |
| **iFCI T** | $PH_2^+$ | -341.6687 | -341.6714 | -341.6722 | -341.6722 | -341.6722 | -341.6722 |
| **iFCI T** | $SiH_2$ | -290.1366 | -290.1389 | -290.1396 | -290.1396 | -290.1396 | -290.1396 |
| **iFCI T** | $C_2H_4$ | -78.2748 | -78.2732 | -78.2718 | -78.2715 | -78.2715 | -78.2715 |
| **iFCI T** | $C_4H_6$ | -155.4210 | -155.4194 | -155.4166 | -155.4160 | -155.4159 | -155.4159 |
| **iFCI T** | Propane | -117.4642 | -117.4546 | -117.4509 | -117.4505 | -117.4506 | -117.4506 |
| **iFCI T** | TMM (triplet) | -155.4731 | -155.4726 | -155.4705 | -155.4702 | -155.4702 | -155.4702 |
| **iFCI T** | Cyclobutadiene | -154.2655 | -154.2314 | -154.2316 | -154.2293 | -154.2350 | -154.2351 |
| **iFCI T** | n-heptane | -274.3148 | -274.2741 | -274.2614 | -274.2566 | -274.2533 | -274.2536 |
| **iFCI T** | 2-methyl-hexane | -274.2519 | -274.2680 | -274.2610 | -274.2565 | -274.2533 | -274.2536 |
| **iFCI T** | m-benzyne | -230.2419 | -230.2341 | -230.2203 | -230.2179 | -230.2177 | -230.2178 |
| **iFCI T** | o-benzyne | -230.2303 | -230.2243 | -230.2130 | -230.2100 | -230.2099 | -230.2100 |
| **iFCI T** | p-benzyne | -230.2589 | -230.2407 | -230.2261 | -230.2238 | -230.2236 | -230.2237 |
| **iFCI T** | $C_6H_{10}$ | -257.0173 | -232.6371 | -232.6300 | -232.6278 | -232.6270 | -232.6272 |
| **iFCI S** | $CH_2$ | -39.0618 | -39.0625 | -39.0625 | -39.0625 | -39.0625 | -39.0625 |
| **iFCI S** | $NH_2^+$ | -55.3394 | -55.3423 | -55.3426 | -55.3426 | -55.3426 | -55.3426 |
| **iFCI S** | $PH_2^+$ | -341.6977 | -341.6996 | -341.6997 | -341.6997 | -341.6997 | -341.6997 |
| **iFCI S** | $SiH_2$ | -290.1700 | -290.1716 | -290.1717 | -290.1717 | -290.1717 | -290.1717 |
| **iFCI S** | $C_2H_4$ | -78.4446 | -78.4432 | -78.4422 | -78.442 | -78.4419 | -78.4419 |
| **iFCI S** | $C_4H_6$ | -155.5455 | -155.5470 | -155.5428 | -155.5418 | -155.5417 | -155.5417 |
| **iFCI S** | Propane | -117.4618 | -117.4516 | -117.4478 | -117.4474 | -117.4476 | -117.4476 |
| **iFCI S** | TMM (singlet) | -155.4527 | -155.4459 | -155.4208 | -155.4382 | -155.4317 | -155.4385 |
| **iFCI S** | Cyclobutadiene | -154.2710 | -154.2407 | -154.2389 | -154.2393 | -154.2423 | -154.2425 |
| **iFCI S** | n-heptane | -274.3134 | -274.2752 | -274.2606 | -274.2551 | -274.2544 | -274.2519 |
| **iFCI S** | 2-methyl-hexane | -274.2522 | -274.2683 | -274.2602 | -274.2551 | -274.2517 | -274.2520 |
| **iFCI S** | m-benzyne | -230.2837 | -230.2639 | -230.2517 | -230.2495 | -230.2495 | -230.2496 |
| **iFCI S** | o-benzyne | -230.2876 | -230.2781 | -230.2674 | -230.2649 | -230.2648 | -230.2651 |
| **iFCI S** | p-benzyne | -230.2517 | -230.2471 | -230.2324 | -230.2301 | -230.2302 | -230.2303 |
| **iFCI S** | $C_6H_{10}$ | -232.7347 | -232.7435 | -232.7352 | -232.7330 | -232.7323 | -232.7322 |

## V. Singlet-Triplet Gap of Copper (II) Acetate Hydrate

Singlet-triplet gaps were computed as vertical transition energies, using the same geometry for each spin state.[2] iFCI is particularly amenable to modeling vertical transitions as the terms in the many-body expansion that do not include the two singly occupied molecular orbitals (SOMOs) cancel exactly. The vertical transition therefore reduces the 43,680 3-body terms of Cu.aqac to only the 2016 terms that involve the SOMOs. The previous investigation using iFCI to model the S-T gap of Cu.aqac used adiabatic transitions and sill only considered the 2016 3-body terms involving at least one SOMO. Due to the operational simplicity of iFCI and iNO-FCI for vertical gaps, we computed vertical gaps for iFCI and iNO-FCI. The geometry for Cu(aqac) can be found in reference 2.

**Table 5.** The energies of each spin state of copper (II) acetate hydrate using iNO-FCI in each basis used in the main manuscript where n refers to the n-body level.

| | **Spin** | **n=1 (Ha)** | **n=2 (Ha)** | **n=3 (Ha)** |
|---|---|---|---|---|
| **6-31g* ($\zeta = 10^{-5.5}$)** | Singlet | -4339.1855 | -4341.2057 | -4341.2365 |
| | Triplet | -4339.1854 | -4341.2057 | -4341.2358 |
| **6-31g* ($\zeta = 10^{-6.5}$)** | Singlet | -4339.1867 | -4341.2184 | -4341.2497 |
| | Triplet | -4339.1866 | -4341.2181 | -4341.2487 |
| **cc-pVTZ(Cu) /6-31g*(other) ($\zeta = 10^{-5.5}$)** | Singlet | -4340.4018 | -4342.4649 | -4342.4832 |
| | Triplet | -4340.4017 | -4342.4644 | -4342.4818 |

## VI. Reaction Mechanism of Criegee Intermediate Reaction with Water

The geometries for this reaction were taken from reference 3.[3]

**Table 6.** The energies of the systems involved in the Criegee intermediate-water reaction in iFCI, 5 DFT functionals and UCCSD(T). TZ represents the cc-pVTZ basis and DZ represents the cc-pVDZ basis.

| | **Criegee** | **Water** | **Intermediate** | **TS** | **Product** |
|---|---|---|---|---|---|
| **iNO-FCI DZ** | -189.1209 | -76.2418 | -265.3786 | -265.3655 | -265.4456 |
| **iNO-FCI TZ** | -189.3155 | -76.3330 | -265.6635 | -265.6499 | -265.7294 |
| **B97-D TZ** | -189.5486 | -76.4165 | -265.9785 | -265.9662 | -266.0284 |
| **PBE TZ** | -188.6720 | -76.0466 | -264.7274 | -264.7086 | -264.7644 |
| **M06-2x TZ** | -189.5686 | -76.4250 | -266.0123 | -266.0041 | -266.0835 |
| **B3LYP TZ** | -189.6566 | -76.4598 | -266.1307 | -266.1204 | -266.1900 |
| **wB97x-D TZ** | -189.5812 | -76.4334 | -266.0306 | -266.0202 | -266.0966 |
| **UCCSD(T) TZ** | -189.3133 | -76.3322 | -265.6604 | -265.6480 | -265.7243 |

## VIII. Geometries

Z-matrix for cis-2-butene:

```
C
C   1 1.513
H   1 1.108  2 110.545
H   1 1.104  2 112.697  3 120.9
H   1 1.108  2 110.541  3 241.8
C   2 XXX  1 127.441  3 239.2
H   6 2.167  2 136.823  1 323.1
H   7 1.789  6  66.449  2  52.2
C   8 1.104  7  36.098  6  36.4
H   9 1.108  8 107.963  7 115.2
H   2 1.102  1 115.369  3  59.2
H   6 1.102  2 117.191  1 180.0
```

Where XXX was replaced with 1.357 for section I and the values in the bond length column in section III

XYZ coordinates of n-octane (taken from NIST CCCBDB database) used for section I:

```
C 4.27089282 -0.20059437 0.00000000
C 3.08295306 0.41436162 0.00000000
C 1.82399044 -0.27861037 0.00000000
C 0.62797682 0.34263638 0.00000000
C -0.62797682 -0.34263638 0.00000000
C -1.82399044 0.27861037 0.00000000
C -3.08295306 -0.41436162 0.00000000
C -4.27089282 0.20059437 0.00000000
H 4.33588926 -1.27748697 0.00000000
H 5.19114767 0.35795720 0.00000000
H 3.04659939 1.49528629 0.00000000
H 1.85100872 -1.36129091 0.00000000
H 0.60473274 1.42551164 0.00000000
H -0.60473274 -1.42551164 0.00000000
H -1.85100872 1.36129091 0.00000000
H -3.04659939 -1.49528629 0.00000000
H -5.19114767 -0.35795720 0.00000000
H -4.33588926 1.27748697 0.00000000
```

XYZ coordinates of n-octane (used in section II)

```
C -0.000 0.764 0.000
C 0.000 -0.764 0.000
C -1.394 1.391 0.000
C 1.394 -1.391 0.000
C -1.394 2.919 0.000
C 1.394 -2.919 0.000
C -2.791 3.531 0.000
C 2.791 -3.531 0.000
H 0.551 1.121 0.877
H 0.551 1.121 -0.877
H -0.551 -1.121 0.877
H -0.551 -1.121 -0.877
H -1.946 1.036 -0.877
```

H -1.946 1.036 0.877
H 1.946 -1.036 -0.877
H 1.946 -1.036 0.877
H -0.840 3.273 0.876
H -0.840 3.273 -0.876
H 0.840 -3.273 0.876
H 0.840 -3.273 -0.876
H -2.760 4.622 0.000
H -3.356 3.218 -0.880
H -3.356 3.218 0.880
H 2.760 -4.622 0.000
H 3.356 -3.218 -0.880
H 3.356 -3.218 0.880

## XYZ coordinates of n-dodecane

C -3.1171853276 3.7191472730 -0.1342927846
C -1.6445176402 3.5057801654 -0.5203083376
H -3.6952677439 2.7942578478 -0.2565540151
H -3.5867804938 4.4919766567 -0.7539238484
H -3.2074934532 4.0301201571 0.9143812422
C -0.9563507332 2.4214400723 0.3307866030
H -1.1030436864 4.4566105073 -0.4207147398
H -1.5793224629 3.2239454528 -1.5820870810
C 0.5092926648 2.1341897069 -0.0562215793
H -1.0016777298 2.7125144481 1.3919832742
H -1.5343222625 1.4889792483 0.2450777948
C 1.4778426138 3.3067355661 0.1912553604
H 0.5520954801 1.8451254583 -1.1180300442
H 0.8615647780 1.2615524115 0.5135777340
C 2.9465338256 2.9553489724 -0.1082736204
H 1.3904731923 3.6289783560 1.2408438896
H 1.1842506394 4.1708597694 -0.4217820357
C 3.9190944939 4.1193644736 0.1546277411
H 3.0376576756 2.6351097077 -1.1578018483
H 3.2435839407 2.0896899334 0.5038099853
C 5.3914653600 3.7662478194 -0.1216890216
H 3.8161445462 4.4456164456 1.2012249405
H 3.6297037093 4.9819032953 -0.4653820862
C 6.3626856593 4.9266306184 0.1621963448
H 5.4995250766 3.4504470188 -1.1708576118
H 5.6766971490 2.8974228321 0.4914758284
C 7.8373165678 4.5717263668 -0.0987060959
H 6.2470369556 5.2460446933 1.2095554633
H 6.0828467470 5.7938998695 -0.4556746439
C 8.8093154939 5.7291259024 0.1957486871
H 7.9565350086 4.2579240837 -1.1473831894
H 8.1156105426 3.7012144293 0.5154632367
C 10.2794615967 5.3644657519 -0.0651524028
H 8.6902163044 6.0427825808 1.2435326676
H 8.5314102708 6.5988208760 -0.4178965634
H 10.5907053212 4.5166949285 0.5583846259
H 10.9466079821 6.2056854978 0.1564885277
H 10.4358779378 5.0806108064 -1.1135823972

## XYZ coordinates of hexadecane

C -11.1411831994 4.4380153815 0.0079169913
C -9.7938201336 3.7030279067 -0.0756929023
H -11.2367635494 4.9784178918 0.9582747296
H -11.9837075448 3.7405996148 -0.0666927133
H -11.2404429666 5.1713921957 -0.8022977810
C -8.5826474401 4.6454538095 0.0438948946
H -9.7325257715 3.1557234894 -1.0280303407
H -9.7408190283 2.9427092888 0.7180294904
C -7.2282305100 3.9206266139 -0.0541821067
H -8.6380344804 5.1861513389 1.0014095951
H -8.6407918688 5.4122187444 -0.7444440448
C -6.0182835429 4.8646351843 0.0614431515

H -7.1755695134 3.3795703506 -1.0115868326
H -7.1692533876 3.1543066535 0.7344422882
C -4.6633857092 4.1426033730 -0.0513545902
H -6.0656686924 5.4000724836 1.0222091697
H -6.0827665497 5.6353855003 -0.7223537641
C -3.4543643182 5.0888860850 0.0550364172
H -4.6187726315 3.6038609122 -1.0104430331
H -4.5954454890 3.3747484538 0.7349591231
C -2.0991251819 4.3688493395 -0.0660103354
H -3.4960562986 5.6248661419 1.0157830862
H -3.5255304576 5.8589386036 -0.7288482674
C -0.8904830533 5.3161238346 0.0360457938
H -2.0587226364 3.8312146067 -1.0259001348
H -2.0264001715 3.6002398241 0.7191256306
C 0.4647722026 4.5961461495 -0.0851685146
H -0.9308391280 5.8536916114 0.9959750934
H -0.9632710560 6.0847898562 -0.7490284185
C 1.6737433636 5.5424866732 0.0212755122
H 0.5064231983 4.0603151620 -1.0459999627
H 0.5360292310 3.8259796686 0.6985970488
C 3.0286890892 4.8206209532 -0.0919998498
H 1.6292565788 6.0809567364 0.9805227344
H 1.6055950447 6.3105604085 -0.7648065087
C 4.2385452185 5.7647340168 0.0237158998
H 3.0759506488 4.2855231164 -1.0529608029
H 3.0934052763 4.0496059643 0.6915190170
C 5.5930387764 5.0401483742 -0.0750724480
H 4.1860700977 6.3053595482 0.9813739806
H 4.1792449385 6.5313988356 -0.7645498084
C 6.8040942216 5.9827101439 0.0446205765
H 5.6482402473 4.4999204909 -1.0328626534
H 5.6515245283 4.2730123739 0.7128818889
C 8.1515548842 5.2479974555 -0.0398186743
H 6.7430121706 6.5295038491 0.9972646648
H 6.7507029597 6.7434394857 -0.7486833148
H 8.2512077724 4.5142002453 0.7699679082
H 8.2469200264 4.7081236458 -0.9904984321
H 8.9939838351 5.9455176137 0.0348915485

## XYZ coordinates of icosane

C -11.2077626142 4.2501966450 0.0092785009
C -9.8459798280 3.5349609560 -0.0619964593
C -8.6468876043 4.4878719265 0.0854283950
H -9.7690236846 2.9994981565 -1.0209704008
H -9.8004462062 2.7644053604 0.7235041868
C -7.2824433000 3.7798719682 0.0010509752
H -8.7212486055 5.0199112858 1.0464063170
H -8.7002402289 5.2608049086 -0.6972067648
C -6.0881764739 4.7435366016 0.1175444616
H -7.2156368639 3.2342397655 -0.9528091179
H -7.2178474824 3.0193332765 0.7946385445
C -4.7195688349 4.0448682982 0.0250389216
H -6.1534967730 5.2899625946 1.0710338273
H -6.1584520645 5.5035715804 -0.6762316525
C -3.5316487520 5.0211931342 0.0976588982
H -4.6647607841 3.4781097475 -0.9172106752
H -4.6337017338 3.3038044859 0.8346935226
C -2.1591760257 4.3307963121 0.0018996057
H -3.5867888669 5.5911039472 1.0379545524
H -3.6212701001 5.7596576780 -0.7141347518
C -0.9764308474 5.3158191097 0.0399827108
H -2.1120928090 3.7446999904 -0.9288777094
H -2.0573719833 3.6079190207 0.8259038873
C 0.3977986768 4.6291595729 -0.0565609659
H -1.0239955760 5.9036690140 0.9696121687
H -1.0796625271 6.0371845088 -0.7852431598
C 1.5800558948 5.6148341430 -0.0195595223
H 0.4451491608 4.0408823204 -0.9859335875

H 0.5015989764 3.9082936573 0.7690172577
C 2.9531037241 4.9257624301 -0.1164606379
H 1.5332618284 6.2010980808 0.9111258325
H 1.4770011244 6.3374771844 -0.8436106434
C 4.1397859943 5.9038045830 -0.0458879518
H 3.0077707051 4.3554252380 -1.0565257901
H 3.0447084641 4.1879378829 0.6956728612
C 5.5093948372 5.2074750950 -0.1410026727
H 4.0857257862 6.4703568941 0.8965281831
H 4.0512382130 6.6448083685 -0.8552922622
C 6.7021001689 6.1731914639 -0.0250869068
H 5.5743193314 4.6622172739 -1.0951841642
H 5.5821934821 4.4467449412 0.6518721660
C 8.0677721889 5.4680794051 -0.1135582290
H 6.6360646407 6.7167989395 0.9299790798
H 6.6344141285 6.9351589445 -0.8170347508
C -12.4011372637 3.2953923272 -0.1505123082
H -11.2483719270 5.0232808619 -0.7726188119
H -11.2867540275 4.7816357220 0.9692543261
H -12.3566264036 2.7679078425 -1.1117642386
H -13.3546408736 3.8347311405 -0.1085981851
H -12.4106173664 2.5373949785 0.6432705049
C 9.2652226820 6.4227848355 0.0356639954
H 8.1419441235 4.9394411637 -1.0764209167
H 8.1236394682 4.6926010976 0.6663893626
C 10.6283129082 5.7105725818 -0.0406425667
H 9.1887036256 6.9542299998 0.9969002353
H 9.2168841761 7.1964151741 -0.7466464715
C 11.8200408351 6.6668815510 0.1223749797
H 10.7072158977 5.1838986733 -1.0032444192
H 10.6713857351 4.9338084750 0.7374773121
H 11.8272110280 7.4287685995 -0.6677050879
H 12.7745137822 6.1295616604 0.0767489015
H 11.7756368436 7.1895866250 1.0862373087